\documentclass[aps,preprint]{revtex4}%
\usepackage{amsfonts}
\usepackage{amsmath}
\usepackage{amssymb}
\usepackage{graphicx}%
\begin{document}
\title[ ]{Electron transport through a coupled double dot molecule: role of inter-dot
coupling, phononic and dissipative effects}
\preprint{ }
\author{M. Imran$^{\ast}$}
\affiliation{Department of Physics, Quaid-i-Azam University, Islamabad, Pakistan.}
\author{B.Tariq}
\affiliation{Department of Physics, Quaid-i-Azam University, Islamabad, Pakistan. National
Center For Physics, Islamabad, Pakistan.}
\author{M. Tahir}
\affiliation{Department of Physics, University of Sargodha, Sargodha, Pakistan}
\author{K. Sabeeh$^{\dagger}$}
\affiliation{Department of Physics, Quaid-i-Azam University, Islamabad, Pakistan}
\pacs{PACS number}

\begin{abstract}
In this work, we have investigated conduction through an artificial molecule
comprising two coupled quantum dots. The question addressed is the role of
inter-dot coupling on electronic transport. We find that the current through
the molecule exhibits step-like features as a function of the voltage between
the leads, where the step size increases as the inter-dot coupling is
increased. These step-like features disappear with increasing tunneling rate
from the leads, but we find that in the presence of coupling, this smooth
behavior is not observed rather two kinks are seen in the current voltage
curve. This shows that the resolution of the two levels persists if there is
finite inter-dot coupling. Furthermore, we also consider the effects of
electron-phonon interaction as well as dissipation on conduction in this
system. Phononic side bands in the differential conductance survive for finite
inter-dot coupling even for strong lead to molecule coupling.

\end{abstract}
\volumeyear{year}
\volumenumber{number}
\issuenumber{number}
\eid{identifier}
\date[Date text]{date}
\received[Received text]{date}

\revised[Revised text]{date}

\accepted[Accepted text]{date}

\published[Published text]{date}

\startpage{1}
\endpage{2}
\maketitle

\subsection{Introduction}

Ever since the original proposal of a single molecule rectifying diode by A.
Aviram and M. A. Ratner\cite{1}, there has been great interest in electron
transport (both charge and spin) at the single molecule level. Theoretical as
well as experimental work in this direction has led to the promising field of
molecular electronics\cite{6}\cite{8}. With an eye on applications it is
expected that the understanding of quantum electron transport at the molecular
scale is a key step to realizing molecular electronic devices\cite{3}.
Experiments on conduction in molecular junctions are becoming more common,
\cite{4}\cite{5} and references therein. On the experimental front, the most
common methods of contacting individual molecules are through scanning
tunneling microscope tips and mechanically controlled break junctions
\cite{6}\cite{8}. Early experiments focused on the absolute conduction and on
trends such as dependence on wire length, molecular structure, and
temperature. From a theoretical point of view, investigating electron
transport in an electrically contacted molecule is a challenging problem. In
this system, the interaction of the electronic degrees of freedom with the
vibrational ones of the molecule need to be considered. In addition, there can
be further complications arising from the electron-electron interaction on the
molecule as well as effects of the environment surrounding the system. Most of
the formal theoretical work on transport in molecular electronics has relied
on the generalized master equations approach \cite{9}\cite{10} and the non
equilibrium Green's function (NEGF) method\cite{7}.

As mentioned above, an important feature that can affect charge transport in a
single molecule is the coupling of electrons to quantized
molecular\ vibrational motion, phonons \cite{19}\cite{20}. For transport
through a single level quantum dot molecule a lot of work has already been
done taking into account vibrational degrees of freedom\cite{11}%
\cite{12}\cite{13}. In this paper, we consider transport through a single
molecule consisting of two coupled quantum dots in parallel configuration.
Electron transport in double quantum dots has been an area of active research,
\cite{34} and references therein. In parallel configuration electron from the
lead can tunnel through either of the two dots. We address the role of finite
coupling between the dots on the electronic transport. We also take into
account the electron-phonon interaction as well as the dissipative effects of
the environment. It has been established that it is important to take into
account electron-phonon interaction in the study of transport in single and
double dot systems\cite{35}. Here we show that the inter-dot coupling will
significantly affect transport when a single electron can occupy either dot in
the presence of electron-phonon interaction. The difference in the transport
properties with and without inter-dot coupling will be discussed in detail in
this work. For a single level molecular system with electron phonon
interaction, phonon side band peaks start disappearing with increasing
tunneling rate\cite{2} whereas we find that for a coupled dot molecule the
phonon peaks survive even if the tunneling rate from the leads is increased.

\subsection{Model}

Our system is a laterally coupled double quantum dot. It is assumed that only
a single level in each dot participates in transport. We allow finite coupling
between the single electronic levels of the two dots. An electron from the
leads can tunnel through either of the two dots.

The full Hamiltonion describing our system is%
\[
H=H_{M}+H_{Leads}+H_{T}.
\]
It is the sum of the electron Hamiltonian of the coupled dot molecule $H_{M}$,
the Hamiltonian of the leads $H_{Leads}$, the tunneling Hamiltonian $H_{T}$
describing the molecule-to-lead coupling. We explain each term in the full
Hamiltonian separately:%
\begin{equation}
H_{M}=\sum_{i=1,\sigma}^{2}\epsilon_{i\sigma}d_{i\sigma}^{\dag}d_{i\sigma
}+\sum_{i,j=1i\neq j,\sigma}^{2}t_{i,j,\sigma}d_{i\sigma}^{\dag}d_{j\sigma}.
\end{equation}
The first term represents two discrete energy levels, one in each dot.
$d_{i\sigma}^{\dag},$ $d_{i\sigma}$ create and annihilate an electron in state
$|i\sigma>$ on the dot$.$ The second term represents inter-dot coupling where
$t_{i,j,\sigma}$ represents coupling between the electronic states of the two
dots. We have assumed $t_{i,j,\sigma}=t_{j,i,\sigma}=t$.%
\begin{equation}
H_{Leads}=\sum_{k}\epsilon_{\nu k\sigma}c_{\nu k\sigma}^{\dag}c_{\nu k\sigma}.
\end{equation}
This represents the leads Hamiltonian. Indices $\nu,k,\sigma$ refer to the
left/right leads, the electronic wave vector in either lead, and leads
electrons spin.

The tunneling Hamiltonian describes hopping between the leads and the
molecule. Direct hopping between the two leads is neglected:%
\begin{equation}
H_{T}=\sum_{i,k}(V_{i,\nu,k,\sigma}c_{\nu k\sigma}^{\dag}d_{i\sigma}+hc).
\end{equation}
The first term represents creation of electrons in the lead and annihilation
of electrons in the coupled dots, while the second term represents creation of
electrons in the coupled dots and annihilation in the lead. Here
$V_{i,\nu,k,\sigma}$ denotes lead-system coupling (hopping) amplitude and $hc$
denotes hermitian conjugation. We consider contacting the coupled dots with
two metallic leads.

Finally, phonons, electron-phonon coupling, heat bath and phonons coupling are
described by the following Hamiltonian\cite{17} (In this calculation we work
with $\hbar=1$):%
\begin{align}
H_{phonon+Bath}  &  =\sum_{q}\omega_{q}a_{q}^{\dag}a_{q}+\sum_{\alpha=1}%
^{2}\sum_{q}\lambda_{\alpha\alpha}^{q}(a_{q}^{\dag}+a_{q})d_{\alpha}^{\dag
}d_{\alpha}+\nonumber\\
&  \sum_{\beta}\omega_{\beta}b_{\beta}^{\dag}b_{\beta}+\sum_{q\beta}N_{q\beta
}(b_{\beta}^{\dag}+b_{\beta})(a_{q}^{\dag}+a_{q}).
\end{align}

The electron-phonon interaction is included with in the first Born
approximation, which is resonable when electron phonons coupling is weak. For
a single dot molecule this problem was studied in \cite{21}\cite{22}, whereas
we consider two dots in the molecule interacting with phonons of frequency
$\omega_{q}$. The first and the third term represents phononic and heat bath
energy. Here $\omega_{q}$, $\omega_{\beta}$ are phonon and heat bath energies.
$a_{q}^{\dag}a_{q}(b_{\beta}^{\dag}b_{\beta})$ are phonons creation and
annihilation operators (heat bath creation and annihilation operators). The
second term represents electron-phonon interaction and $\lambda_{\alpha\alpha
}^{q}$ is the coupling strength of this interaction. The last term represents
phonon and heat bath coupling and $N_{q\beta}$ is the coupling strength of
phonon heat bath interaction.

\subsection{Method}

Our approach is based on the nonequilibrium Green function technique\cite{29}%
\cite{30}, which is now a standard technique in mesoscopic physics as well as
molecular electronics. We follow the formulation pioneered by Meir and
Wingreen\cite{31}, Jauho and co-workers\cite{14}\cite{33}. The case of
intermediate and strong electron-phonon coupling at finite tunneling rates is
the most interesting regime but it is also the most difficult. Only the
approaches by Flensberg\cite{24}. and Galperin et al\cite{17}exist, both
starting from the exact solution for the isolated system and then switching on
tunneling as a perturbation\cite{2}.The current from lead $\nu$ is given by
the well known expression\cite{32}%
\begin{align}
J^{\nu}(\epsilon)  &  =\frac{e}{2\pi}\sum_{i,j,k}\int_{-\infty}^{\infty
}d\epsilon\lbrack\{G_{ij}^{r}(\epsilon)-G_{ij}^{a}(\epsilon)\}\Sigma_{ji\nu
k}^{<}(\epsilon)+\\
&  G_{ij}^{<}(\epsilon)\{\Sigma_{ji\nu k}^{a}(\epsilon)-\Sigma_{ji\nu k}%
^{r}(\epsilon)\}].\nonumber
\end{align}
Here $\Sigma_{ji\nu k}^{<,r,a}$ represents lesser, retarded and advanced self
energies of leads and coupled dot molecule. $\Sigma_{ji,\nu}^{<,r,a}=\sum
_{k}\{V_{j\nu k}^{\dag}g_{\nu}^{<,r,a}(\epsilon)V_{i\nu k}\}$ where $g_{\nu
}^{<,r,a}(\epsilon)$ is the lesser,retarded and advanced Green's function of
the leads.\cite{31}.

We employ the wide-band approximation, where the self-energy of the coupled
dot molecule due to each lead is taken to be energy independent and is given
by%
\[
\sum_{k}\Sigma_{ji\nu k}^{r,a}(\epsilon)=DV_{i\nu k}V_{j\nu k}^{\dag}%
\int_{-\infty}^{\infty}\dfrac{d\epsilon_{k}}{(\epsilon-\epsilon_{k}\pm i\eta
)}\qquad\eta\rightarrow0^{\mp}=-i2\pi D(\epsilon_{\nu})V_{i\nu k}V_{j\nu
k}^{\dag}=\mp i\dfrac{\Gamma_{ji}^{\nu}}{2}.
\]
Here $D$ is the constant energy density of the leads. Similiarly the lesser
self energy can be written as%
\begin{equation}
\sum_{k}\Sigma_{ji\nu k}^{<}(\epsilon)=i\Gamma_{ji}^{\nu}f(\epsilon)
\end{equation}
where $\Gamma_{ji}^{\nu,}s$ are the tunneling rates (coupling of leads with
the molecule) and $f(\epsilon)$ is the Fermi-Dirac distribution function.

Now by employing the current symmetrization and line width proportionality
approximations\cite{14} we obtain%
\begin{equation}
J^{\nu}(\epsilon)=\frac{ie}{2\pi}\sum_{i,j}\left(  \frac{\Gamma_{ji}^{L}%
\Gamma_{ji}^{R}}{\Gamma_{ji}^{L}+\Gamma_{ji}^{R}}\right)  \int_{-\infty
}^{\infty}d\epsilon\lbrack\{G_{ij}^{r}(\epsilon)-G_{ij}^{a}(\epsilon
)\}\{f^{L}(\epsilon)-f^{R}(\epsilon)\}].
\end{equation}
By using the equation of motion technique, we work out the coupled dot
molecule Green's function and find%
\begin{equation}
(g_{ii}^{r}(\epsilon))^{-1}=\left(  \epsilon-\epsilon_{i}-\frac{\mid
t_{i,j}\mid^{2}}{\epsilon-\epsilon_{j}}\right)  \qquad i\neq j.
\end{equation}%
\begin{equation}
(g_{i,j}^{r}(\epsilon))^{-1}=\frac{\left(  \epsilon-\epsilon_{j}\right)
}{t_{j,i}}\left(  \epsilon-\epsilon_{i}-\frac{\mid t_{i,j}\mid^{2}}%
{\epsilon-\epsilon_{j}}\right)  \qquad i\neq j
\end{equation}
Employing Dyson's equation, we can write%
\begin{equation}
G_{ij}^{r}(\epsilon)=\left[  (g_{ij}^{r}(\epsilon))^{-1}-\Sigma_{ji\nu k}%
^{r}(\epsilon)\right]  ^{-1}.
\end{equation}
The differential conductance in the absence of electron-phonon interaction is%
\begin{equation}
\frac{dJ^{\nu}}{dV}=\frac{e^{2}}{2\pi}\sum_{i,j}\left[  \dfrac{\Gamma_{ji}%
^{L}\Gamma_{ji}^{R}}{\left(  (g_{ij}^{r}(\epsilon))^{-1}\right)  ^{2}+\left(
\dfrac{\Gamma_{ji}^{L}+\Gamma_{ji}^{R}}{2}\right)  ^{2}}\right]  .
\end{equation}
If the electron-phonon interaction is included then our total self energy will
become%
\[
\Sigma_{Total}=\Sigma_{Leads}+\Sigma_{el+phonon}.
\]
and in the Hartree-Fock approximation%
\begin{equation}
\Sigma_{el+ph}=\Sigma_{Hartree}+\Sigma_{Fock}%
\end{equation}
where%
\begin{equation}
\Sigma_{ji,Hartree}^{r}(\epsilon)=-i\sum_{q}\lambda_{q}^{2}\int_{-\infty
}^{\infty}\frac{d\epsilon}{2\pi}D_{0}^{r}(q,\epsilon=0)G_{ji}^{<}(\epsilon)
\end{equation}
and%
\begin{align}
\Sigma_{ji,Fock}^{r}(\epsilon)  &  =-i\sum_{q}\lambda_{q}^{2}\int_{-\infty
}^{\infty}\frac{d\epsilon^{\prime}}{2\pi}\left[  D_{0}^{r}(q,\epsilon
-\epsilon^{\prime})G_{ji}^{<}(\epsilon^{\prime})+\right. \nonumber\\
&  \left.  D_{0}^{r}(q,\epsilon-\epsilon^{\prime})G_{ji}^{r}(\epsilon^{\prime
})+D_{0}^{<}(q,\epsilon-\epsilon^{\prime})G_{ji}^{r}(\epsilon^{\prime
})\right]  .
\end{align}
Here $D_{0}^{r,<}$ represents retarded and lesser free phonon Green's
function. At this stage if we also include the effects of the heat bath then
our zeroth order phonon Green's function will be%
\begin{equation}
D_{0}^{r}(\epsilon,\omega_{q})=\dfrac{1}{\epsilon-\omega_{q}+\dfrac{i\gamma
}{2}}-\dfrac{1}{\epsilon+\omega_{q}+\dfrac{i\gamma}{2}}%
\end{equation}
with $\gamma=2\pi\sum_{\beta}|N_{q\beta}|^{2}\delta(\epsilon-\omega_{\beta})$,
\cite{17}, represents dissipation in phonon energy due to its contact with the
heat bath.%
\begin{equation}
D_{0}^{<}(\epsilon)=-in(\epsilon)\dfrac{\gamma}{(\epsilon-\omega_{q}%
)^{2}+\left(  \dfrac{\gamma}{2}\right)  ^{2}}-i(1+n(\epsilon))\dfrac{\gamma
}{(\epsilon+\omega_{q})^{2}+\left(  \dfrac{\gamma}{2}\right)  ^{2}}%
\end{equation}
with $n(\epsilon)$ the Bose-Einstein distribution function. At zero
temperature the imaginary part of the electron-phonon self energy is%
\begin{equation}
\text{Im}\Sigma_{ji,el+phonon}=\frac{1}{2}\sum_{q}\lambda_{q}^{2}%
[A_{ji}(z-\omega_{q})-A_{ji}(z+\omega_{q})]
\end{equation}
where the spectral function $A_{ji}(\epsilon)=-2\operatorname{Im}G_{ji}%
^{r}(\epsilon),$ whereas
\begin{equation}
\operatorname{Re}\Sigma_{ji,el+phonon}=\sum_{q}\lambda_{q}^{2}\left[
\operatorname{Re}G_{ji}^{r}(z-\omega_{q})-\operatorname{Re}G_{ji}^{r}%
(z+\omega_{q})+\dfrac{1}{2\omega_{q}}\right]
\end{equation}
with
\begin{equation}
z\longrightarrow\epsilon+\frac{i\gamma}{2}.
\end{equation}
In the presence of electron-phonon interaction and the dissipative effects of
the heat bath, the differential current is%
\begin{equation}
\frac{dJ^{\nu}}{dV}=\frac{e^{2}}{2\pi}\sum_{i,j}\dfrac{\Gamma_{ji}^{L}%
\Gamma_{ji}^{R}}{\Gamma_{ji}^{L}+\Gamma_{ji}^{R}}\dfrac{\left(  \Gamma
_{ji}^{L}+\Gamma_{ji}^{R}+Im\Sigma_{ji,el+phonon}\right)  }{\left(
(g_{ij}^{r}(\epsilon))^{-1}+\operatorname{Re}\Sigma_{ji,el+phonon}\right)
^{2}+\left(
\genfrac{}{}{1pt}{0}{\Gamma_{ji}^{L}+\Gamma_{ji}^{R}}{2}%
+Im\Sigma_{ji,el+phonon}\right)  ^{2}}%
\end{equation}

\subsection{Results And Discussions}

The system under consideration is a molecule comprising two coupled quantum
dots. This molecule is attached to two metallic leads. The primary focus of
this study is the role of inter-dot coupling in electronic transport. In
addition, the dissipative effects of the environment is taken into account by
treating the transport in this system surrounded by a heat bath of phonons.
The vibrational states of the molecule and its impact on electronic transport
is included through the electron-phonon interaction. Our results show
characteristic non-ohmic behavior in the current-voltage results presented in
Fig.(1). At this stage, we are ignoring the phononic and heat bath effects in
order to highlight the role of inter-dot coupling. When coupling of the system
to the leads, which enters through the tunneling-rate, is very small then
energy levels of the two dots are sharply peaked. Hence, no current flows
until the applied voltage bias is in resonance with the level of either of the
two dots. We see in Fig.(1) that intially no current flows on increasing the
bias voltage. But as the applied bias comes in resonance with the level of the
first dot a sharp increase in current occurs. Further increase in applied bias
does not lead to increase in current because the level of the second dot is
not in resonance. As the applied bias is further increased, it comes in
resonance with the level of the second dot leading to an abrupt increase in
current. This explains the step-like features with sharp steps and plateaus
observed in Fig.(1). Now we consider the effects of inter-dot coupling. As a
result of the coupling, the levels of the two dots are pushed apart. This
results in the plateaus becoming wider as this requires higher bias voltage
before the Fermi level in the lead is in resonance with the higher level of
the double dot molecule. This is also shown in Fig.(1) as the inter-dot
coupling is increased. Instead of the inter-dot coupling, if we increase the
tunneling rate from the lead, broadening of the electronic states in the two
dots of the molecule takes place. In this situation, current increases
linearly with applied bias and the system exhibits ohmic behavior. Now if we
increase the inter-dot coupling, step like features again begin to appear as
the applied voltage is tuned since the coupling pushes the two levels apart.
For sufficiently large tunneling rate the two states broaden to the extent
that they merge and we find that the current increases smoothly with
increasing applied bias without any step-like features, Fig.(2).%

\begin{figure}
[ptb]
\begin{center}
\includegraphics[
height=3.3287in,
width=5.3195in
]%
{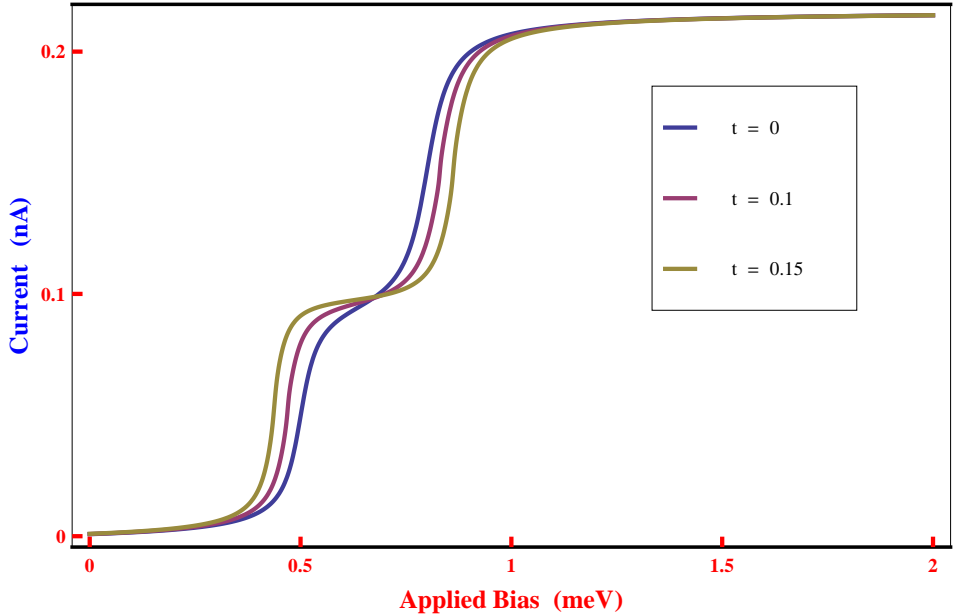}%
\caption{Current as a function of applied bias. The energies of two levels are
$\epsilon_{_{1}}=0.5meV$, $\epsilon_{_{2}}=0.8meV$ . The tunneling rates from
the two states are $\Gamma_{ii}=0.04meV$, $\Gamma_{ij}=0meV,$ $where$ $i\neq
j.$}%
\end{center}
\end{figure}
%

\begin{figure}
[ptb]
\begin{center}
\includegraphics[
height=3.3287in,
width=5.3195in
]%
{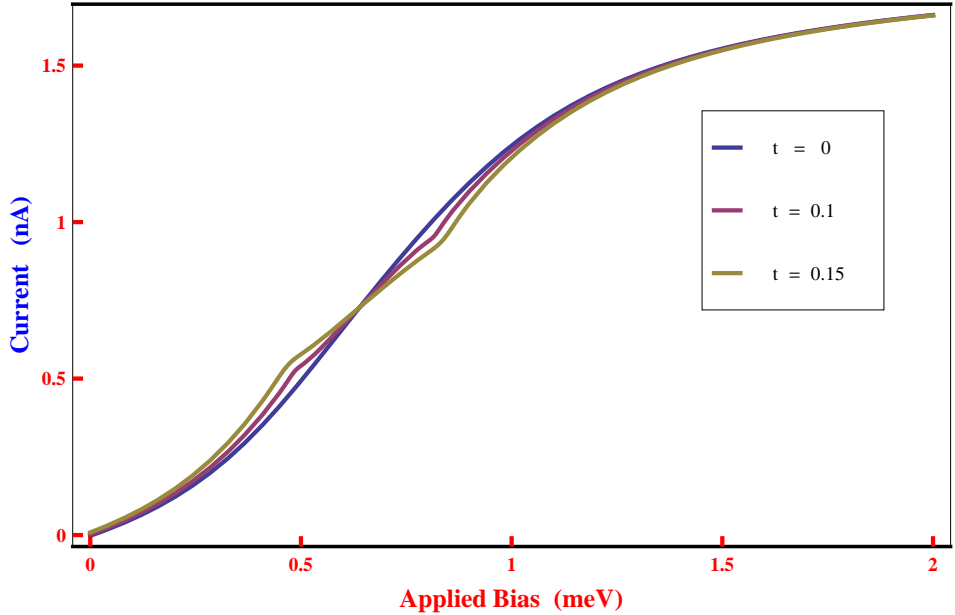}%
\caption{Current as a function of applied bias. The energies of two levels are
$\epsilon_{_{1}}=0.5meV$, $\epsilon_{_{2}}=0.8meV$ . The tunneling rates from
the two states are $\Gamma_{ii}=0.4meV$, $\Gamma_{ij}=0meV,$ $where$ $i\neq
j.$}%
\end{center}
\end{figure}

These results are also presented in Fig(3) where it is seen that if the lead
to the system coupling is small then the energy states of the two dots are
sharp and this feature appears as two peaks in the differential conductance.
As the lead to the system coupling is increased, the electronic states get
broadened. And for sufficiently large tunneling rate (strong lead to system
coupling) both the states merge and the peaks in the differential conductance
disappear. To observe the effects of inter-dot coupling with in the molecule
even as the tunneling rate from the lead to the molecule (lead to molecule
coupling) is increased, we show in Fig.(4) that for finite inter-dot coupling,
the peaks in the differential conductance persist. The electronic states are
broadened due to the coupling of the leads and the molecule but when inter-dot
coupling is taken into account, the difference in energy between the levels
increases. This compensates the broadening of the levels and allows the two
levels to remain distinct. This results in peaks corresponding to the two
levels appearing in differential conductance inspite of broadening of the
levels, Fig.(4).

\bigskip%
\begin{figure}
[ptb]
\begin{center}
\includegraphics[
height=3.3287in,
width=5.3195in
]%
{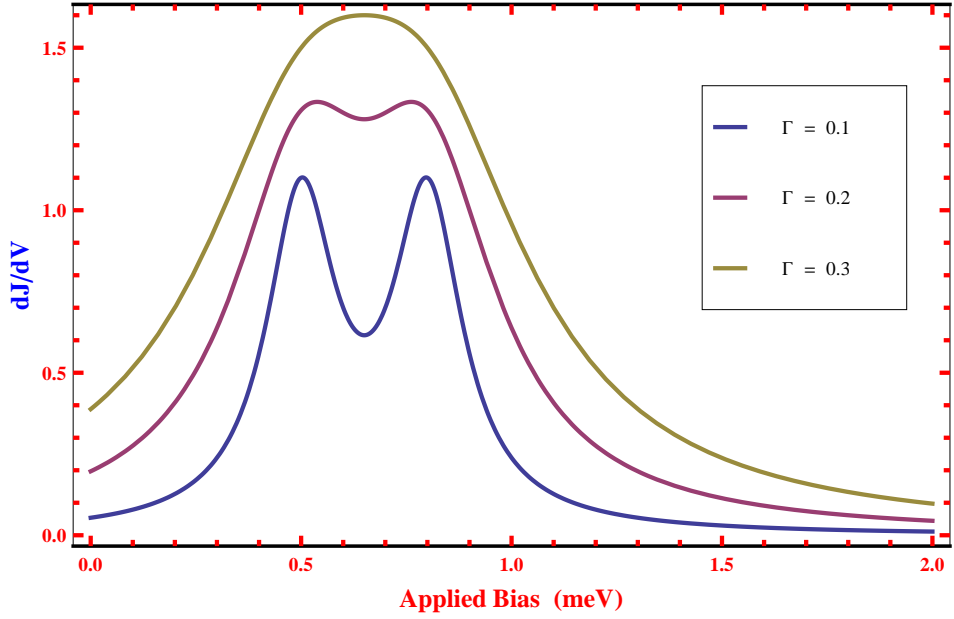}%
\caption{Differential conductance $(dJ/dV)$ as a function of applied bias and
tunneling rates. Applied bias and tunneling rates are in the units of meV. The
energies of two levels are $\epsilon_{1}=0.5meV$ ,$\epsilon_{2}=0.8meV$
,$t=0meV$.\ }%
\end{center}
\end{figure}
\begin{figure}
[ptb]
\begin{center}
\includegraphics[
height=3.3287in,
width=5.3195in
]%
{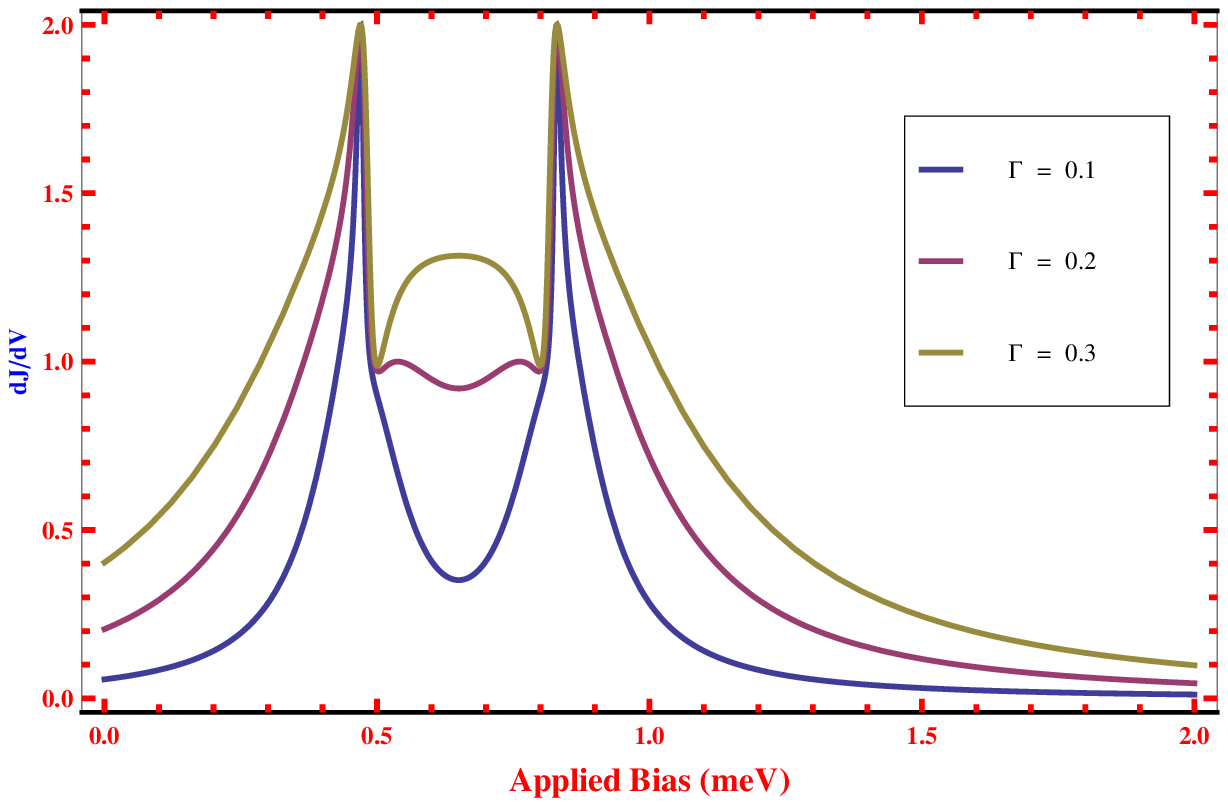}%
\caption{Differential conductance $(dJ/dV)$ as a function of applied bias and
tunneling rates. Applied bias and tunneling rates are in the units of meV. The
energies of two levels are $\epsilon_{1}=0.5meV$ ,$\epsilon_{2}=0.8meV$
,$t=0.1meV$.\ }%
\end{center}
\end{figure}

\bigskip

At the final stage, we consider the effects of the electron-phonon
interaction. These are shown in Figs (5) and (6). We find that in addition to
peaks in the differential conductance corresponding to the two levels of the
dots there are peaks due to phonons. These phononic peaks (side bands) occur
as the electrons can exchange energy with the phonons and contribute to
conductance. To focus on the role of inter-dot coupling on phononic peaks, we
see that as we increase the (lead to molecule coupling) tunneling rate
electronic states are broadened to the extent that phononic side bands are not
visible. On further increase in tunneling rate the two electronic states
broaden and merge into each other, Fig.(5). If we include inter-dot coupling,
not only the electronic states remain distinct but the phononic effects are
not lost either. This can be seen in Fig.(6) where peaks appear corresponding
to the two electronic states as well as the phononic side band peaks. Even
with an increase in tunneling rate from the leads to the molecule, these
features persist in the presence of inter-dot coupling.

\bigskip%
\begin{figure}
[ptb]
\begin{center}
\includegraphics[
height=3.6236in,
width=5.3195in
]%
{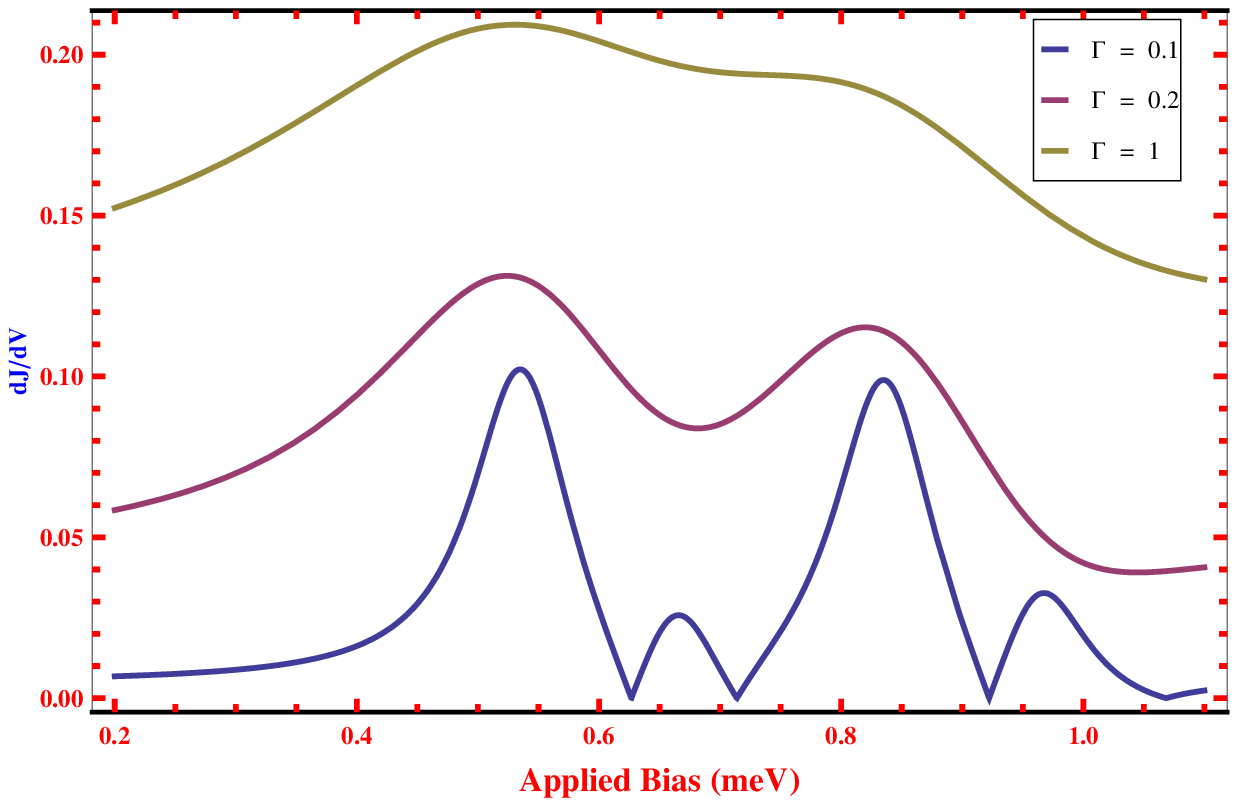}%
\caption{Differential conductance $(dJ/dV)$ as a function of applied bias and
tunneling rates. Applied bias and tunneling rates are in the units of meV. The
energies of two levels are $\epsilon_{1}=0.5meV$ ,$\epsilon_{2}=0.8meV$
,$t=0meV$ ,$\ \omega=0.1meV$ ,$\gamma=0.013meV$.\ }%
\end{center}
\end{figure}
%

\begin{figure}
[ptb]
\begin{center}
\includegraphics[
height=2.2157in,
width=3.1981in
]%
{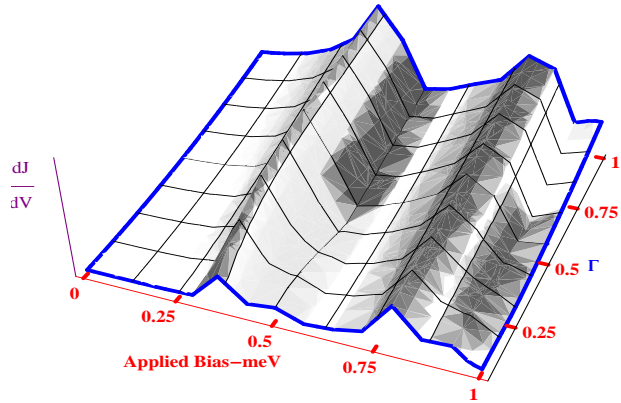}%
\caption{Differential conductance $(dJ/dV)$ as a function of applied bias and
tunneling rates. Applied bias and tunneling rates are in the units of meV. The
energies of two levels are $\epsilon_{1}=0.5meV$ ,$\epsilon_{2}=0.8meV$
,$t=0.1meV$ ,$\ \omega=0.1meV$ ,$\gamma=0.013meV$.\ }%
\end{center}
\end{figure}

\bigskip

\bigskip

\bigskip

\bigskip

\bigskip

\bigskip

\bigskip

\bigskip

\bigskip

\bigskip

\bigskip

\bigskip

\bigskip

\bigskip

\bigskip

\bigskip

\bigskip

\bigskip

\bigskip

\bigskip

\bigskip

\bigskip

\bigskip

\bigskip

\bigskip

\bigskip

\bigskip

\bigskip

\bigskip

\bigskip

To conclude, in this work we have focused on the role of inter-dot coupling
with in the dot molecule on electron transport. We have considered a coupled
dot molecule, with inter-dot coupling, attached to two leads including
electron-phonon interaction and the coupling of the molecule with an
environment allowing dissipation of the phonons. We find that including
inter-dot coupling has profound and important role in transport. The step like
featues in current voltage characteristics of peaks in differential
conductance corresponding to the the dot energy levels is lost in the absence
of inter-dot coupling when strong coupling to the leads is considered.
Inter-dot coupling allows the two levels to remain distinct with peaks
appearing in the differential conductance even when broadening of the levels
in the dots occur for strong lead to molecule coupling. Furthermore, phononic
side bands that appear in the differential conductance also persist in the
presence of finite inter-dot coupling even for strong lead to molecule coupling.

\bigskip

\subsubsection{Acknowledgements}

M. Imran and K. Sabeeh would like to acknowledge the support of the Higher
Education Commission (HEC) of Pakistan through project No. 20-1484/R\&D/09.

$^{^{\ast}}$imran1gee@gmail.com

$^{\dagger}$kashifsabeeh@hotmail.com


\bigskip

\begin{thebibliography}{99}                                                                                               %


\bibitem {1}A. Aviram, M.A. Ratner, Chem. Phys. Lett.\textbf{ 29}, 277 (1974)

\bibitem {2}M. A. Reed and J. M. Tour, Sci. Am. 282 , \textbf{86}.(2000).

\bibitem {3}G. Cuniberti, G. Fagas, and K. Richter, Introducing Molecular
Electronics, Springer-Verlag, Berlin, (2005) .

\bibitem {4}Molecular Nanoelectronics, edited by M. A. Reed and T. LeeAmerican
Scientific Publishers, Stevenson Ranch, CA, 2003 .

\bibitem {5}D. R. Bowler, J. Phys.: Condens. Matter \textbf{16}, R721(2004) .

\bibitem {6}M.A. Reed, C. Zhou, C.J. Muller, T.P. Burgin, J.M. Tour, Science
\textbf{278}, 252 (1997)

\bibitem {7}L. V. Keldysh, Zh. Eksp. Teor. Fiz. \textbf{47}, 1515 1965 ; H.
Huag and A. P. Jauho, Quantum Kinetics in Transport and Optics of
Semiconductors, Springer Solid-State Sciences Vol.\textbf{ 123 }Springer, New
York, (1996).

\bibitem {8}E. L%
\"{}%
ortscher, H.B. Weber, H. Riel, Phys. Rev. Lett. \textbf{98}, 176807 (2007)

\bibitem {9}U. Weiss, Quantum Dissipative Systems, Series in Modern Condensed
Matter Physics, vol. \textbf{10 }(World Scientific, 1999)

\bibitem {10}H.P. Breuer, F. Petruccione, The theory of open quantum systems
(Oxford University Press, Oxford, 2002)

\bibitem {11}I.G. Lang, Y.A. Firsov, Sov. Phys. JETP \textbf{16}, 1301 (1963)

\bibitem {12}A.C. Hewson, D.M. Newns, Japan. J. Appl. Phys. Suppl. 2, Pt. 2,
121 (1974)

\bibitem {13}G. Mahan, Many-Particle Physics, 2nd edn. (Plenum, New York, 1990)

\bibitem {14}Antti-Pekka Jauho, Ned S. Wingreen, and Yigal Meir Phys. Rev. B
\textbf{50}, 5528 (1994)

\bibitem {15}Qing-feng Sun and X. C. Xie, Phys.Rev B\textbf{ 75}, 155306 (2007).

\bibitem {16}A. Hewson and D. Newns, J. Phys. C \textbf{13}, 4477 (1980)

\bibitem {17}Michael Galperin, Abraham Nitzan,and Mark A. Ratner Phys. Rev. B
\textbf{73}, 045314 (2006)

\bibitem {18}Z.-Z. Chen, R. L\"{u}, and B.-F. Zhu, Phys. Rev. B \textbf{71}, 165324(2005).

\bibitem {19}M. Di Ventra, S.T. Pantelides, and N. D. Lang, Phys. Rev.Lett.
\textbf{88}, 046801 (2002).

\bibitem {20}T. Seideman, J. Phys. Condens. Matter \textbf{15}, R521 (2003).

\bibitem {21}D. A. Ryndyk and J. Keller, Phys. Rev. B\textbf{ 71}, 073305 (2005).

\bibitem {22}D. A. Ryndyk, M. Hartung, and G. Cuniberti, Phys. Rev. B
\textbf{73},045420 (2006) .

\bibitem {23}S. Datta, W. Tian, S. Hong, R. Reifenberger, J. I. Henderson,
andC. P. Kubiak, Phys. Rev. Lett.\textbf{ 79}, 2530 (1997) .

\bibitem {24}K. Flensberg, Phys. Rev. B \textbf{68}, 205323,(2003).

\bibitem {25}T. Frederiksen, M. Brandbyge, N. Lorente, and A.-P. Jauho, Phys.
Rev. Lett. \textbf{93}, 256601 (2004) .

\bibitem {26}M. Galperin, M. A. Ratner, and A. Nitzan, Nano Lett.\textbf{ 4},
1605 (2004) .

\bibitem {27}M. Galperin, M. A. Ratner, and A. Nitzan, J. Phys. Chem.
\textbf{121},11965 (2004) .

\bibitem {28}M. Galperin, M. A. Ratner, and A. Nitzan, J. Phys.:
Condens.Matter \textbf{19}, 103201 (2007) .

\bibitem {29}L. Kadanoff and G. Baym, Quantum Statistical Mechanics
Benjamin,New York, (1962).

\bibitem {30}J. Rammer and H. Smith, Rev. Mod. Phys. \textbf{58}, 323 (1986).

\bibitem {31}Y. Meir and N. S. Wingreen, Phys. Rev. Lett. \textbf{68}, 2512 (1992).

\bibitem {32}M. Tahir and A. MacKinnon Phys.Rev B\textbf{ 77}, 224305 (2008).

\bibitem {33}A.-P. Jauho, J. Phys.: Conf. Ser. \textbf{35}, 313 (2006).

\bibitem {34}W.G. van der Wiel, \textit{et.al, }Rev. Mod. Phys. \textbf{75}, 1 (2003).

\bibitem {35}S.Tarucha, T. Fujisawa, K. Ono, D.G. Austin, T.H. Oosterkamp, and
W.G. van der Wiel, Microelectron. Eng. \textbf{47}, 101(1999); T. Fujisawa,
T.H.Oosterkamp, W.G. van der Wiel, B.W. Broer, R. Aguado, S. Tarucha, and L.P.
Kouwenhoven, Science \textbf{282}, 932(1998); H. Qin, F. Simmel, R.H. Blick,
J.P. Kotthaus, W. Wegscheider, and M. Bichler, Phys. Rev. B \textbf{63},
035320 (2001); T. Brandes and B. Kramer, Phys. Rev. Lett. \textbf{83}, 3021
(1999); T. Brandes and N. Lambert, Phys. Rev. B \textbf{67}, 125323 (2003); T.
Fujisawa, D.G. Austing, Y. Tokura, Y. Hirayama, and S. Tarucha, Nature
(London) \textbf{419}, 278 (2002).
\end{thebibliography}
\end{document}